\documentclass[11pt]{article}

\usepackage{graphics}
\usepackage{a4wide}
\usepackage{amsfonts}

\newcommand{\nit}{\noindent}

\newcommand{\np}{\newpage}
\newcommand{\dsp}{\displaystyle}
\newcommand{\vs}[1]{\vspace{#1 ex}}
\newcommand{\hs}[1]{\hspace{#1 em}}
\newcommand{\bflr}{\begin{flushright}}
\newcommand{\eflr}{\end{flushright}}
\newcommand{\bc}{\begin{center}}
\newcommand{\ec}{\end{center}}
\newcommand{\ben}{\begin{enumerate}}
\newcommand{\een}{\end{enumerate}}

\newcommand{\be}{\begin{equation}}
\newcommand{\ee}{\end{equation}}
\newcommand{\ba}{\begin{array}}
\newcommand{\ea}{\end{array}}
\newcommand{\ct}{\cite}
\newcommand{\bit}{\bibitem}
\newcommand{\dd}[2]{\frac{\partial{#1}}{\partial{#2}}}

\newcommand{\bg}{\beta}

\newcommand{\del}{\delta}
\newcommand{\eps}{\epsilon}
\newcommand{\ve}{\varepsilon}

\newcommand{\thg}{\theta}

\newcommand{\lb}{\lambda}

\newcommand{\rg}{\rho}

\newcommand{\vf}{\varphi}

\newcommand{\Gam}{\Gamma}

\newcommand{\Og}{\Omega}
\newcommand{\Lb}{\Lambda}

\newcommand{\bfm}{{\bf m}}
\newcommand{\bfn}{{\bf n}}
\newcommand{\bfp}{{\bf p}}
\newcommand{\bfr}{{\bf r}}
\newcommand{\bfs}{{\bf s}}
\newcommand{\bft}{{\bf t}}
\newcommand{\bfv}{{\bf v}}
\newcommand{\bfA}{{\bf A}}
\newcommand{\bfB}{{\bf B}}
\newcommand{\bfC}{{\bf C}}
\newcommand{\bfF}{{\bf F}}
\newcommand{\bfJ}{{\bf J}}
\newcommand{\bfK}{{\bf K}}
\newcommand{\bfL}{{\bf L}}

\newcommand{\bfrg}{\mbox{{\boldmath $\rg$}}}

\newcommand{\bfPi}{\mbox{{\boldmath $\Pi$}}}
\newcommand{\bfnab}{\mbox{\boldmath $\nabla$}}

\newcommand{\cD}{{\cal D}}

\newcommand{\cQ}{{\cal Q}}

\newcommand{\lh}{\left(}
\newcommand{\rh}{\right)}

\newcommand{\der}{\partial}

\begin{document}

\pagestyle{empty} 
\vs{3}
\begin{center} 
{\Large{\bf Covariant hamiltonian dynamics}} \\
\vs{7} 

{\large J.W.\ van Holten }\\
\vs{2} 

{\large NIKHEF, Amsterdam NL$^*$}\\ 
\vs{3} 

\today 
\vs{15} 

{\small{\bf Abstract}}
\end{center} 

\nit
{\footnotesize{We discuss the covariant formulation of the dynamics 
of particles with abelian and non-abelian gauge charges in external fields.
Using this formulation we develop an algorithm for the construction of 
constants of motion, which makes use of a generalization of the 
concept of Killing vectors and tensors in differential geometry. 
We apply the formalism to the motion of classical charges in abelian 
and non-abelian monopole fields. }} 
\vfill
\footnoterule 
\nit
{\footnotesize{$^*$\tt{e-mail: v.holten@nikhef.nl}}}
\np 
~\hfill

\np

\pagestyle{plain}
\pagenumbering{arabic} 

\section{Introduction}

A standard problem in the solution of classical and quantum mechanical
systems is to identify the constants of motion associated with the 
system. There are many ways to obtain constants of motion, but two
methods are fairly common in different branches of physics. In the 
canonical phase-space formulation of classical conservative systems  
we have to identify all dynamical quantities $Q$ which commute with 
the hamiltonian in the sense of Poisson brackets\footnote{In quantum 
mechanics, the Poisson bracket is replaced by the commutator; for
brevity and to avoid irrelevant operator ordering complications, we
stay with classical mechanics in this paper.}:
\be 
\left\{ Q, H \right\} = 0.
\label{0.1}
\ee
The draw-back of this prescription is, that for systems with gauge
interactions this formulation is usually not manifestly gauge covariant. 

On the other hand for systems with a non-flat configuration space, 
such as particles moving on a curved manifold (or space-time, in 
general relativity) the appropriate algorithm is to search for Killing 
vectors and their higher-rank generalizations. In Riemannian geometry 
these are covariant objects, but the procedure is only applicable for
geodesic motion in the absence of non-geometrical external fields of 
force. 

As observed in \ct{jackiw-manton}, for constants of motion to exist 
in the case of non-geodesic motion, e.g., for particles in external 
fields, the symmetries of the metric and those of the external fields 
have to match. In fact Killing vectors appear explictly in the 
expressions for constants of motion linear in the momentum. In 
\ct{rh,grh} a complete set of consistency conditions for the existence 
of constants of motion were derived for particles in arbitrary 
background geometries, using a covariant hamiltonian phase-space 
approach including the contributions of spin. This procedure also 
applied to constants of motion which are higher-order polynomials in 
the momentum, as well as constants of motion which are Grassmann-odd 
expressions in the spin degrees of freedom, which generate standard
or non-standard supersymmetries on the worldline of the system.  

In this paper we show how to extend this covariant phase-space 
approach to include the presence of external gauge fields. As 
in \ct{jackiw-manton} our non-abelian dynamics is based on the 
equations of motion postulated by Wong \ct{wong}. These equations
were studied in a geometric setting, using the method of co-adjoint 
orbits, for a similar purpose in \ct{duval-horv}, whilst a lagrangean 
realization in terms of Grassmann variables was constructed in 
\ct{linmfvh}. Having a completely covariant phase-space formulation 
we derive a set of generalized Killing equations, the solution of 
which produces all constants of motion in a manifestly covariant way. 
To avoid unnecessary complications, we formulate all our dynamical 
models in euclidean or riemannian space, but the generalization to 
minkowskian or lorentzian manifolds is straightforward.

\section{Dynamics of point charges}

The classical dynamics of a point charge in a magnetic field is described 
by the Lorentz force law
\be
m \dot{\bfv} = q \bfv \times \bfB
\label{1.1}
\ee
In the standard canonical formulation this equation is derived from 
a hamiltonian
\be 
H = \frac{1}{2m} \lh \bfp - q \bfA \rh^2,
\label{1.2}
\ee
via Hamilton's equations:
\be 
\dot{\bfr} = \dd{H}{\bfp} = \frac{1}{m} \lh \bfp - q \bfA \rh, \hs{2}
\dot{\bfp} = - \dd{H}{\bfr} = \frac{q}{m} \lh \bfnab \bfA \rh \cdot 
\lh \bfp - q \bfA \rh 
\label{1.3}
\ee
Therefore 
\be
\bfp = m \dot{\bfr} + q \bfA = m \bfv + q \bfA, \hs{2}
\dot{\bfp} = q\, \bfnab \bfA \cdot \bfv,
\label{1.4}
\ee
and after substitution and the definition $\bfB = \bfnab \times \bfA$
eq.\ (\ref{1.1}) follows. In terms of the field-strength tensor $\bfF$ 
\be
 F_{ij} = \ve_{ijk} B_k = \nabla_i A_j - \nabla_j A_i,
\label{1.5}
\ee
the equation for the Lorentz force takes the form
\be
m \dot{v}_i = q F_{ij} v_j.
\label{1.6}
\ee
In this form the equation can be extended easily to relativistic
particles in Minkowski space. 

The above construction uses cartesian co-ordinates $r_i$ and their 
canonical momenta $p_i$, such that the equations of motion an be
written in terms of Poisson brackets
\be
\left\{ f, g \right\} = \sum_i \dd{f}{r_i} \dd{g}{p_i}
 - \dd{f}{p_i} \dd{g}{r_i},
\label{1.7}
\ee
for phase-space functions $f(r_i,p_i,t)$ and $g(r_i, p_i,t)$. 
In term of these brackets
\be 
\frac{df}{dt} = \dd{f}{t} + \left\{ f, H \right\}.
\label{1.8}
\ee
The phase-space co-ordinates $(r_i, p_i)$ are canonical as
the only non-trivial fundamental bracket is 
\be 
\left\{ r_i, p_j \right\} = \del_{ij},
\label{1.9}
\ee
all others vanishing:
\be 
\left\{ r_i, r_j \right\} = \left\{ p_i, p_j \right\} = 0.
\label{1.10}
\ee
A disadvantage of this formulation is, that the canonical momenta
are gauge dependent: 
\be 
\bfA^{\prime} = \bfA + \bfnab \Lb \hs{1} \Rightarrow \hs{1}
\bfp^{\prime} = \bfp + q \bfnab \Lb,
\label{1.11}
\ee
although this does not affect the fundamental brackets (\ref{1.9}). 
As a result, the hamiltoninan equations of motion are not manifestly
gauge covariant. 

However, an alternative exists in which the dynamical variables of the
particle are all gauge invariant, and which has the added advantage that 
the hamiltonian takes a very simple form. Introduce the gauge-invariant
momenta
\be
\bfPi = \bfp - q \bfA = m \bfv.
\label{1.12}
\ee
Then the hamiltonian takes the simple quadratic form 
\be 
H = \frac{1}{2m}\, \bfPi^2.
\label{1.13}
\ee
This has the form of a free-particle hamiltonian, but the dynamics
is now manifest in the modified brackets:
\be 
\left\{ f, g \right\} = \dd{f}{r_i} \dd{g}{\Pi_i} 
 - \dd{f}{\Pi_i} \dd{g}{r_i} + q F_{ij} \dd{f}{\Pi_i} \dd{g}{\Pi_j}. 
\label{1.14}
\ee
In particular the fundamental brackets are
\be 
\left\{ r_i, \Pi_j \right\} = \del_{ij}, \hs{2} 
\left\{ r_i, r_j \right\} = 0, \hs{2}
\left\{ \Pi_i, \Pi_j \right\} = - q\, F_{ij}.
\label{1.15}
\ee
This shows, that the momenta $\bfPi$ are not canonical, but act like
covariant derivatives, rather than ordinary partial derivatives; 
indeed, the last bracket is the Poisson-bracket version of the 
Ricci identity. As a result, we can derive the homogeneous Maxwell
equations (the Bianchi identities) from the Jacobi identity:
\be 
\left\{ \Pi_i, \left\{ \Pi_j, \Pi_k \right\} \right\}
 + \left\{ \Pi_j, \left\{ \Pi_k, \Pi_i \right\} \right\}
 + \left\{ \Pi_k, \left\{ \Pi_i, \Pi_j \right\} \right\} = 0 
\label{1.16.1}
\ee
which implies
\be
\nabla_i F_{jk} + \nabla_j F_{ki} + \nabla_k F_{ij} = 0 
 \hs{1} \Leftrightarrow \hs{1} \bfnab \cdot \bfB = 0.
\label{1.16}
\ee
It remains to establish that the brackets and the hamiltonian reproduce
the correct equations of motion; this follows by direct computation:
\be 
\dot{r}_i = \left\{ r_i, H \right\} = \frac{\Pi_i}{m}, \hs{2}
\dot{\Pi}_i = \left\{ \Pi_i, H \right\} = \frac{q}{m}\, F_{ij} \Pi_j.
\label{1.17}
\ee
The covariant phase-space formulation has been used by various authors 
in different contexts, see e.g.\ \ct{jackiw-manton,grh,duval-horv}.

\section{Symmetries and constants of motion}

The gauge-covariant formulation of hamiltonian mechanics of
charged particles is mathematically elegant; it is also most suited 
to study the existence of symmetries and constants of motion.  
Indeed, in the hamiltonian framework a constant of motion 
$Q(\bfr, \bfPi)$ is identified by the property that its bracket
with the hamiltonian vanishes:
\be
\left\{ Q, H \right\} = 0 \hs{1} \Rightarrow \hs{1}
 \Pi_i \lh \nabla_i Q - q F_{ij} \dd{Q}{\Pi_j} \rh = 
 \bfPi  \cdot \lh \bfnab Q + q \bfB \times \dd{Q}{\bfPi} \rh = 0. 
\label{2.1}
\ee
A systematic procedure is to expand any constant of motion 
as a power series in $\bfPi$:
\be 
Q(\bfr, \bfPi) = C(\bfr) + C_i(\bfr) \Pi_i 
 + \frac{1}{2} C_{ij}(\bfr) \Pi_i \Pi_j + ...\; .
\label{2.10}
\ee
Substitution gives a series of constraints
\be 
\ba{l}
\dsp{ \nabla_i C = q F_{ij} C_j , }\\
 \\
\dsp{ \nabla_i C_j + \nabla_j C_i = q F_{ik} C_{kj} 
 + q F_{jk} C_{ki}, }\\
 \\
\dsp{ \nabla_{i} C_{jk} + \nabla_j C_{ki} + \nabla_k C_{ij} 
 = q F_{il} C_{ljk} + q F_{jl} C_{lki} + q F_{kl} C_{lij}, }\\
 \\
...
\ea 
\label{2.12}
\ee
This series can be truncated whenever there is a Killing vector or 
tensor of flat space:
\be 
\nabla_{(i_1} C_{i_2 ... i_n)} = 0.
\label{2.13}
\ee
Then we can take $C_{i_1 ... i_p} = 0$ for all $p \geq n$, and the 
constant of motion takes the polynomial form
\be 
Q(\bfr, \bfPi) = \sum_{k = 0}^{p-1}\, \frac{1}{k!}\, C_{i_1 ... i_k}(\bfr)\, 
 \Pi_{i_1} ... \Pi_{i_k}.
\label{2.13.1}
\ee 
Note that it is always possible to add an arbitary constant to the zeroth
order coefficient $C(\bfr)$. Therefore it is obvious that for particles in
an electromagnetic background there are no non-trivial constants of motion 
corresponding to only a $C(\bfr)$ with $C_i(\bfr)$, $C_{ij}(\bfr)$ and all 
higher co-efficients vanishing. The first non-trivial case is therefore the 
set truncated at $p = 2$:
\be 
Q (\bfr, \bfPi) = C(\bfr) + \bfC (\bfr) \cdot \bfPi,
\label{2.13.4}
\ee
with $\bfC$ a Killing vector of flat space:
\be 
\nabla_i C_j + \nabla_j C_i = 0.
\label{2.13.5}
\ee
Such Killing vectors generate translations and rotations, and take the form
\be 
\bfC = \bfm + \bfn \times \bfr,
\label{2.13.6}
\ee
where $\bfm$ and $\bfn$ are arbitrary constant vectors. 
The lowest-order constraint equation now becomes
\be 
\nabla_i C = q F_{ij} C_j =
 q \nabla_i \lh A_j C_j \rh - q A_j \nabla_i C_j - q C_j \nabla_j A_i
\label{2.13.7}
\ee
Now use the Killing equation for $C_i$ to rewrite this equation as
\be 
\nabla_i \lh C - q \bfC \cdot \bfA \rh = q \bfA \cdot \bfnab C_i
 - q \bfC \cdot \bfnab A_i. 
\label{2.13.8}
\ee
A very simple example a constant magnetic field:
\be 
\bfA = \frac{\bg}{2}\, \bfC \hs{1} \Rightarrow \hs{1}
\bfB = \bg \bfn.
\label{2.13.9}
\ee
For such a field
\be 
C = q\, \bfC \cdot \bfA = \frac{q \bg}{2}\, \bfC^2 
 \hs{1} \mbox{(mod constant)}.
\label{2.13.10}
\ee
As a result the full constant of motion takes the form
\be 
\ba{lll}
Q & = & \bfC \cdot \lh q \bfA + \bfPi \rh = \bfC \cdot \bfp  \\
 & & \\
 & = & \dsp{ \bfm \cdot \bfp 
 + \frac{1}{\bg}\, \bfB \cdot \lh \bfr \times \bfp \rh. }
\ea
\label{2.13.11}
\ee
As $\bfm$ is arbitrary, all components of the momentum and the component 
of the angular momentum in the direction of $\bfB$ are conserved. This 
reflects the invariance under translations and transverse rotations 
in a constant magnetic field. 

Starting from the general Killing vector (\ref{2.13.6}) the equation 
for the Killing scalar becomes
\be 
\bfnab C = q\, \bfm \times \bfB + q \lh \bfr\, \bfn \cdot \bfB 
 - \bfn\, \bfr \cdot \bfB \rh.
\label{2.14.0}
\ee
As a non-trivial example, consider axially symmetric fields 
$\bfB$ with the axis defined by a unit vector $\bfn$:
\be
\bfB = \frac{B(\rg)}{\rg}\, \bfr \times \bfn 
 = \frac{B(\rg)}{\rg}\, \bfrg \times \bfn, \hs{2}
\bfrg = \bfr - \lh \bfr \cdot \bfn \rh \bfn, 
\label{2.14.1}
\ee 
where $\rg = |\bfrg|$. Then equation(\ref{2.14.0}) becomes
\be
\bfnab C = q \bfm \times \bfB = \frac{qB}{\rg}\, 
 \lh \bfrg\, \bfn \cdot \bfm - \bfn\, \bfrg \cdot \bfm \rh.
\label{2.14.2}
\ee
The corresponding vector potential can be taken as
\be 
\bfA = g(\rg) \bfn,
\label{2.14}
\ee 
provided we identify $B(\rg) = g^{\prime}(\rg)$.
Then eq.\ (\ref{2.14.2}) becomes
\be 
\bfnab C = \frac{q g^{\prime}(\rg)}{\rg}\, \lh \bfrg\, \bfn \cdot \bfm
 - \bfn\, \bfrg \cdot \bfm \rh = 
q \lh \bfm \cdot \bfn \rh \bfnab g - q \lh \bfm \cdot \bfnab g \rh \bfn.
\label{2.15.1}
\ee 
This equation allows solutions for $\bfm = \lb \bfn$, with $\lb$ an
arbitrary scalar parameter. As $\bfn \cdot \bfrg = 0$ and $\bfn^2 = 1$, 
one finds
\be 
C = \lb q\, g(\rg).
\label{2.15.2}
\ee
The full constant of motion then reads
\be 
\ba{lll}
Q & = & \dsp{ 
 \lb q\, g(\rg) + \lh \lb \bfn + \bfn \times \bfr \rh \cdot \bfPi }\\
 & & \\
 & = & \dsp{ \lb \bfn \cdot \bfp + \bfn \cdot \lh \bfr \times \bfp \rh. } 
\ea
\label{2.15.3}
\ee
As $\lb$ is arbitrary, it follows that the components of the canonical
momentum and the angular momentum in the direction $\bfn$ are 
independently conserved\footnote{Actually, the components of the 
canonical and covariant angular momentum in the direction $\bfn$ are 
the same.} 

One can also search for constants of motion which are higher-order 
polynomials in the momentum, associated with flat-space Killing tensors. 
The simplest one is 
\be 
C_{ij} = \del_{ij}, 
\label{2.23}
\ee
which has the special property that
\be 
F_{ij}C_{jk} + F_{kj} C_{ji} = 0.
\label{2.23.1}
\ee
Therefore the associated Killing vector and scalar can be taken to vanish,
and the corresponding constant of motion is the hamiltonian:
\be 
C = \frac{1}{2}\, \del_{ij} \Pi_i \Pi_j = mH,
\label{2.24}
\ee
More complicated Killing tensors are of the form
\be 
C_{ij} = 2\, \del_{ij} \bfn \cdot \bfr - (n_i r_j + n_j r_i),
\label{2.25}
\ee
for an arbitrary fixed unit vector $\bfn$, and
\be 
C_{ij} = \del_{ij} \bfr^2 - r_i r_j,
\label{2.26}
\ee
which is the radial counterpart of (\ref{2.25}). 
Any constants of motion associated with these Killing tensors
are extensions of the $\bfn$-component of the Runge-Lenz vector
\be 
\bfn \cdot \bfK = \bfn \cdot \lh \bfPi \times \bfL \rh
 = \bfn \cdot \bfr\, \bfPi^2 - \bfn \cdot \bfPi\, \bfr \cdot \bfPi,
\label{2.27}
\ee
or the total angular momentum
\be 
\bfL^2 = \lh \bfr \times \bfPi \rh^2 
 = \bfr^2\, \bfPi^2 - \lh \bfr \cdot \bfPi \rh^2 = \bfr \cdot \bfK.
\label{2.28}
\ee
Such constants of motion are associated with special field configurations,
in particular spherically symmetric ones. We discuss such fields in the 
next section.

\section{Magnetic monopoles}

A spherically symmetric magnetic solution of the Maxwell equations is the 
Dirac monopole:
\be
\bfB = \frac{g\bfr}{r^3}.
\label{2.18}
\ee
For such a field equation (\ref{2.14.0}) takes the form 
\be 
\bfnab C = \frac{qg}{r^3} \lh \bfm \times \bfr 
 + (\bfr \cdot \bfn) \bfr - r^2\, \bfn \rh.
\label{2.20}
\ee
Now the first term is a curl, not a gradient; as a result we have to 
take $\bfm = 0$, and
\be 
C = - qg\, \frac{\bfn \cdot \bfr}{r}.
\label{2.20.1}
\ee
The result for the constants of motion based on the
Killing vector (\ref{2.13.6}) then is
\be 
Q = \bfn \cdot \lh - gq\, \frac{\bfr}{r} + \bfr \times \bfPi \rh,
\label{2.21}
\ee
where $\bfn$ is an arbitrary vector; therefore all components of
the gauge-covariant improved angular momentum 
\be
\bfJ = \bfL - q \bfr \times \bfA - gq\, \frac{\bfr}{r}, \hs{2} 
\bfL = \bfr \times \bfp,
\label{2.22}
\ee
are conserved. We observe, that these quantities generate rotations
and satisfy the standard $so(3)$ Lie algebra
\be 
\left\{ J_i , J_j \right\} = \ve_{ijk} J_k.
\label{2.22.1}
\ee 
The Casimir invariant of $so(3)$ is the total angular momentum squared:
\be 
\bfJ^2 = \bfL^2 + g^2 q^2,
\label{2.29}
\ee
which is a constant of motion with
\be 
C_{ij} = \del_{ij} \bfr^2 - r_i r_j, \hs{2} 
C_i = 0, \hs{2} C = g^2 q^2.
\label{2.30}
\ee
It follows from (\ref{2.29}), that the values of the total angular
momentum in the classical theory satisfy the bound $\bfJ^2 \geq C = 
g^2 q^2$. A recent analysis relating many different formulations of 
this dynamicalsystem is found in \ct{plyushchay}. 

Remarkably, the Runge-Lenz vector can be extended to a constant 
of motion in another type of central magnetic field:
\be 
\bfB = \frac{g\bfr}{r^{5/2}}.
\label{2.31}
\ee
Indeed, in such a field there is a constant of motion
\be 
Q = \bfn \cdot \lh \bfK + \frac{2gq}{\sqrt{r}}\, \bfL 
 - 2 g^2 q^2\, \frac{\bfr}{r} \rh,   
\label{2.32}
\ee
for any unit vector $\bfn$, with
\be
\ba{l} 
\dsp{ C_{ij} = 2\, \del_{ij}\, \bfn \cdot \bfr - (n_i r_j + n_j r_i), }\\ 
 \\
\dsp{ C_i = \frac{2gq}{\sqrt{r}}\, \bfn \times \bfr, }\\
 \\
\dsp{ C = - 2g^2 q^2\, \frac{\bfn \cdot \bfr}{r}. }
\ea
\label{2.33}
\ee
In contrast, for such a central field there is no conserved angular 
momentum vector $\bfJ$, although the total angular momentum $\bfJ^2$ 
is conserved. Of course, the total energy of the magnetic field 
(\ref{2.31}) diverges both at $r = 0$ and at $r \rightarrow \infty$, 
and the field does not satisfy the free Maxwell equations. Therefore it 
requires non-trivial magnetic sources and boundary conditions. However, 
even if the field would exist only in a restricted region of space, the 
constant of motion exists provided the orbit of the point charge is 
also restricted to this region.

\section{Non-abelian point charges}

The gauge-covariant dynamics of point charges can be extended to 
non-abelian point charges. The starting point is defined by the
Wong equations \ct{wong}, which can be written in the form
\be 
\ba{l}
\bfPi = m \bfv, \hs{2} \dot{\bfPi} = g\, t_a\, \bfv \times \bfB_a, \\
 \\
\dot{t}_a = g f_{abc}\, \bfv \cdot \bfA_b\, t_c,
\ea
\label{3.1}
\ee
where the $t_a$ are the non-abelian gauge variables, and where
the non-abelian field strength is defined as
\be 
F_{ij\,a} = \ve_{ijk}\, B_{k\,a} = \nabla_i A_{j\,a} - \nabla_j A_{i\,a}
 - g f_{abc}\, A_{i\,b} A_{j\,c}.
\label{3.1.1}
\ee
A lagrangean representation of the gauge variables in terms of 
Grassmann-odd degrees of freedom was shown to exist in ref.\ \ct{linmfvh}. 
The hamiltonian formulation uses canonical co-ordinates and momenta, 
with a hamiltonian
\be 
H = \frac{1}{2m} \lh \bfp - g \bfA_a t_a \rh^2,
\label{3.2}
\ee
supplemented by the fundamental brackets
\be 
\left\{ r_i, p_j \right\} = \del_{ij}, \hs{2}
\left\{ t_a, t_b \right\} = - f_{abc}\, t_c,
\label{3.3}
\ee
with all other brackets vanishing. It is easily verified, that these
brackets reproduce the equations of motion (\ref{3.1}). A more transparent 
and gauge-covariant formulation is obtained by introduction of the 
covariant momentum
\be 
\bfPi = m \bfv = \bfp - g \bfA_a t_a.
\label{3.4}
\ee
with the quasi-free hamiltonian
\be 
H = \frac{1}{2m}\, \bfPi^2.
\label{3.5}
\ee
The equations of motion (\ref{3.1}) are now reobtained from
the covariant brackets
\be 
\dot{f} = \left\{ f, H \right\},
\label{3.5.1}
\ee
where the brackets are defined explicitly by 
\be 
\left\{ f, h \right\} = \cD_i f\, \dd{h}{\Pi_i} 
 - \dd{f}{\Pi_i}\, \cD_i h + g F_{ij\,a} t_a\, 
 \dd{f}{\Pi_i} \dd{h}{\Pi_j} - f_{abc} t_c \dd{f}{t_a} \dd{h}{t_b}. 
\label{3.6}
\ee
Here the covariant phase-space derivative of a function $f(\bfr, \bfPi, t_a)$ 
appearing on the right-hand side is defined as
\be 
\cD_i f = \nabla_i f - g f_{abc} t_a A_{i\,b} \dd{f}{t_c}.
\label{3.7}
\ee 
As in the abelian case we can now look for constants of motion by
applying the hamiltonian formalism:
\be 
\left\{ Q, H \right\} = 0 \hs{1} \Rightarrow \hs{1}
\Pi_i \lh \cD_i Q - g F_{ij\,a} t_a\, \dd{Q}{\Pi_j} \rh = 0,
\label{4.1}
\ee
or in vector notation
\be 
\bfPi \cdot \lh \mbox{\boldmath{$\nabla$}} Q - g t_a f_{abc} \bfA_b 
 \dd{Q}{t_c} + g t_a \bfB_a \times \dd{Q}{\bfPi} \rh = 0.
\label{4.2}
\ee
Using a covariant momentum expansion
\be 
Q = C + C_i \Pi_i + \frac{1}{2}\, C_{ij} \Pi_i \Pi_j + ...,
\label{4.3}
\ee
we obtain a set of constraints to be satisfied 
\be 
\ba{l}
\cD_i C = g t_a F_{ij\,a} C_j, \\
 \\
\cD_i C_j + \cD_j C_i = g t_a \lh F_{ik\,a} C_{kj} + g F_{jk\,a} C_{ki} \rh, \\
 \\
\cD_i C_{jk} + \cD_j C_{ki} + \cD_k C_{ij} =
 g t_a \lh F_{il\,a} C_{ljk} + F_{jl\,a} C_{lki} + F_{kl\,a} C_{lij} \rh, \\
 \\
...
\ea
\label{4.4}
\ee
Clearly the condition for the series (\ref{4.3}) to stop after a finite 
number of terms is the existence of a tensor satisfying the condition
\be 
\cD_{(i_1} C_{i_2 ...i_n)} = 0,
\label{4.5}
\ee
a gauge-covariant generalization of the Killing equation. All of this
is a direct non-abelian generalization of the case of Maxwell-Lorentz 
theory. 

\section{2-D Yang-Mills theory}

A simple example to illustrate the procedure of constructing constants
of motion from generalized Killing-vectors and tensors is offered by
2-$D$ $SU(2)$ Yang-Mills theory. In this theory the magnetic field strength 
is represented by a triplet of scalar fields $B^{a}$:
\be
F_{ij}^{a} = B^{a} \ve_{ij},
\label{5.1}
\ee
which satisfies the euclidean Yang-Mills equations
\be 
D_i B^{a} = 0.
\label{5.2}
\ee
Thus the  magnetic field is covariantly constant, and its modulus
$B^2 = B^{a} B^{a}$ is a gauge-invariant real constant:
\be 
\nabla_i B^2 = 0.
\label{5.3}
\ee
As the effect of a local gauge transformation is a local rotation of
$B^{a}$ in the internal space, it is possible to gauge transform 
the magnetic field locally into a constant, e.g.
\be 
B^{a} = (0, 0, B).
\label{5.4}
\ee
Such a constant magnetic field can be constructed from the linear
gauge potential
\be 
A_{i}^{a} = - \frac{1}{2}\, B^{a} \ve_{ij} r_j.
\label{5.4.1}
\ee 
In a constant magnetic field there is translation and
rotation invariance, hence we can look for Killing vectors
of the type (\ref{2.13.6}):
\be 
C_i = m_i - \lb\, \ve_{ij} r_j,
\label{5.5}
\ee
with $m_i$ and $\lb$ arbitrary constants. It then only remains
to solve for the generalized Killing scalar $C = C^{a} t_a$:
\be 
\ba{lll}
\lh \cD_i C \rh^{a} & = & 
 \nabla_i C^{a} - g \eps^{abc} A_{i}^{b} C^{c} \\
 & & \\
 & = &  g F_{ij}^{a} C_j\, =\, gB^{a} \lh \ve_{ij} m_j + \lb r_i \rh. 
\ea
\label{5.6}
\ee
The straightforward solution to this equation with $A_i^{a}$ given by
(\ref{5.4.1}) is
\be 
C^{a} = g B^{a} \lh \ve_{ij} r_i m_j + \frac{\lb}{2}\, \bfr^2 \rh.
\label{5.7}
\ee
With $m_i$ and $\lb$ we thus associate two constants of
motion; a gauge-improved momentum:
\be 
Q_i = \Pi_i - g B^{a} t_a \ve_{ij} r_j = p_i + g A_i^{a} t_a,
\label{5.8}
\ee
and the canonical angular momentum
\be
J = \ve_{ij} r_i \Pi_j + \frac{g}{2}\, B^{a} t_a \bfr^2
 = \ve_{ij} r_i p_j.
\label{5.9}
\ee
As a next step, one can look for symmetric Killing tensors $C_{ij}$;
however, the two candidates
\be 
C_{ij} = \del_{ij}, \hs{2} C_{ij} = \del_{ij} \bfr^2 - r_i r_j,
\label{5.10}
\ee
lead us back to the hamiltonian (\ref{3.5}) and the square
of the angular momentum $J^2$, respectively. A non-trivial
constant of motion of this type is a Runge-Lenz-like vector
\be 
\ba{lll}
K_i & = & \dsp{ 
 r_i\, \bfPi^2 - \Pi_i\, \bfPi \cdot \bfr + g B^{a} t_a
 \lh \frac{1}{2}\, \ve_{ij} \Pi_j \bfr^2 + r_i \ve_{jk} r_k \Pi_k \rh
 + \frac{1}{2}\, \lh g B^{a}t_a \rh^2 r_i\, \bfr^2 }\\
 & & \\
 & = & \dsp{ r_i\, \bfp^2 - p_i\, \bfp \cdot \bfr + \frac{1}{4}\, 
 g B^{a} t_a\, J\, r_i - \frac{1}{8}\, \lh g B^{a} t_a \rh^2 r_i\, \bfr^2. }
\ea
\label{5.11}
\ee 
These constants of motion, described simultaneously by the arbitrary
linear combination $\bfn \cdot \bfK$, are constructed in terms of the 
generalized Killing tensors
\be 
\ba{l}
\dsp{ C_{ij} = 2 \del_{ij}\, \bfr \cdot \bfn - r_i n_j - r_j n_i, }\\
 \\
\dsp{ C_i = - g B^{a} t_a\, \ve_{ij} \lh r_j \bfr \cdot \bfn
 + \frac{1}{2}\, n_j \bfr^2 \rh, }\\
 \\
\dsp{ C = \frac{1}{2}\, \lh g B^{a} t_a \rh^2 \bfr^2\, \bfr \cdot \bfn. }
\ea
\label{5.12}
\ee
This solution of the 2-$D$ Yang-Mills equations can be embedded 
straightforwardly in 3-$D$ Yang-Mills theory by taking connections
\be 
A_x^{a} = - \frac{1}{2}\, B^{a} y, \hs{2} A_y^{a} = \frac{1}{2}\, B^{a} x,
\label{6.1}
\ee
whilst all other components of $A_{\mu}^{a}$ vanish. Actually, this amounts 
simply to embedding an abelian solution of the type (\ref{2.13.9}) into a 
non-abelian model. In essence, the results of the abelian case 
are reproduced.

\section{The non-abelian Wu-Yang monopole}
 
A genuinely non-abelian static solution of the pure $SU(2)$ 
Yang-Mills equations in 3-$D$ space is the non-abelian Wu-Yang  
monopole \ct{wu-yang}; the monopole field is given by 
\be 
A_i^{a} = \frac{1}{g} \frac{\ve_{iak} r_k}{r^2}, 
\label{6.0}
\ee
with the corresponding magnetic field strength
\be
F_{ij}^{a} = \frac{1}{g} \frac{\ve_{ijk} r_k r_a}{r^4}, \hs{2}
D_j F_{ij}^{a} = 0.
\label{6.0.1}
\ee
In such a field a particle has a conserved charge invariant under 
combined spatial and isospin rotations
\be 
\cQ = \frac{r^{a} t_a}{r} \hs{1} \Rightarrow \hs{1} 
\cD_i \cQ = 0.
\label{6.0.2}
\ee
In addition, there is a conserved angular momentum
\be 
\bfn \cdot \bfJ = \bfn \cdot \lh \bfr \times \bfPi 
 - \cQ \frac{\bfr}{r} \rh \hs{1} \Leftrightarrow \hs{1}
 \bfJ = \bfr \times \bfp - \bft = \bfL - \bft.
\label{6.0.3}
\ee
which is constructed from the Killing vector and associated scalar
\be 
C_i = \lh \bfn \times \bfr \rh_i, \hs{2}
C = - \cQ\, \frac{\bfn \cdot \bfr}{r} 
 = - \bfn \cdot \bfr\, \frac{r^{a} t_a}{r^2}.
\label{6.0.4}
\ee 
It is easily established that the components of the angular momentum 
$\bfJ$ generate the $so(3)$ Lie algebra (\ref{2.22.1}). The 
contribution of isospin to the orbital angular momentum mixes gauge 
and spin degrees of freedom, a result well-known in the literature 
\ct{jackiw-rebbi,hasenfratz-thooft}. For point-particles carrying 
even-dimensional isospin representations (e.g., isodoublets) 
this turns the bound states into fermions \ct{goldhaber}. 

Also in this case there exist constants of motion quadratic in
the momenta: the hamiltonian $H$ and the square of the total angular 
momentum:
\be
\bfJ^2 = \bfr^2 \bfPi^2 - \lh \bfr \cdot \bfPi \rh^2 
 + \lh \frac{r^{a} t_a}{r} \rh^2.
\label{6.0.5}
\ee
This constant of motion is constructed from the Killing tensor 
(\ref{2.26}):
\[
C_{ij} = \del_{ij} \bfr^2 - r_i r_j,
\]
with $C_i = 0$ and $C = \cQ^2$. In contrast, a constant of motion 
polynomial in the momenta which generalizes the Runge-Lenz vector 
does not exist for a point-particle in the Wu-Yang monopole background.

As is well-known, there also exist abelian monopole solutions in 
spontaneously broken non-abelian gauge theories \ct{thooft,polyakov}. 
The motion of non-abelian point particles in such a background has 
been studied in \ct{wipf,feher}.

\section{Charged particles in curved space}

The concept of Killing vector has its origin in differential 
geometry, where it arises as generator of an isometry. In the 
previous sections we have applied the concept in flat space, the 
isometries of which are translations and rotations. We have shown 
how the concept can be generalized in the presence of background 
gauge fields, abelian as well as non-abelian. Symmetries and constants 
of motion arise in particular when the isometries are matched by
symmetries of the background fields. We have also discussed 
Killing tensors of higher rank, associated with constants of 
motion depending on higher powers of the momenta. 

The generalizations can easily be extended to non-flat spaces. 
The hamiltonian of a charged particle moving in a space with metric 
$g_{ij}(x)$ is
\be 
H = \frac{1}{2m}\, g^{ij}(x)\, \Pi_i \Pi_j.
\label{7.1}
\ee
For a particle without spin the covariant brackets 
are the same as in flat space:
\be 
\left\{ x^{i}, \Pi_j \right\} = \del^{i}_j, \hs{2}
\left\{ \Pi_i, \Pi_j \right\} = q F_{ij}.
\label{7.2}
\ee
In particular, the equations of motion become
\be 
\ba{l}
\dsp{ \dot{x}^{i} = \left\{ x^{i}, H \right\} 
 = \frac{1}{m}\, g^{ij} \Pi_j, \hs{1} \Leftrightarrow \hs{1}
 \Pi_i = m g_{ij}\, \dot{x}^{j}, }\\
 \\
\dsp{ \dot{\Pi} = \left\{ \Pi_i, H \right\} = 
 \frac{1}{m}\, g^{kl} \lh \Gam_{ik}^{\;\;\;j} \Pi_l \Pi_j
 + q F_{ik} \Pi_l \rh \hs{1} \Leftrightarrow \hs{1}
\frac{D \Pi_i}{Dt} = \dot{\Pi}_i - \dot{x}^k\, \Gam_{ki}^{\;\;\;j}
 \Pi_j = q F_{ij}\, \dot{x}^j. }
\ea
\label{7.3}
\ee
By the first of these equations, the second one reduces to the 
Lorentz-Wong equations in curved space. As before, constants of 
motion are obtained by solving the equation
\be
\left\{ Q, H \right\} = 0,
\label{7.4.1}
\ee
with $Q$ a polynomial in the momenta
\be
Q(x,\Pi) = C(x) + C^{i} (x) \Pi_i + \frac{1}{2}\, C^{ij}(x) \Pi_i \Pi_j
 + ...
\label{7.4.2}
\ee
Then the coefficients are solutions of the hierarchy of differential
equations
\be
\ba{l}
D_i C = \nabla_i C = q F_{ij} C^{j}, \\
 \\
D_i C_j + D_j C_i = q F_{ik} C^{k}_{\;\;j} + q F_{jk} C^{k}_{\;\;i}, \\
 \\
D_i C_{jk} + D_j C_{ki} + D_k C_{ij} = q F_{il} C^{l}_{\;\;jk}
 + q F_{jl} C^{l}_{\;\;ki} + q F_{kl} C^{l}_{\;\;ij}, \\
 \\
 ...
\ea
\label{7.4.3}
\ee
As usual, indices are raised and lowered with the metric, and the 
covariant derivative $D_i$ is constructed with the Levi-Civita 
connection
\be 
D_i C_j = \nabla_i C_j - \Gam_{ij}^{\;\;\;k} C_k,
\label{7.4.4}
\ee 
in the case of abelian background gauge fields. 
In the case of non-abelian background gauge fields
we have to make the replacements
\be 
D_i\, \rightarrow\, \cD_i = D_i - g f_{abc} t_a A_{i\,b} \dd{}{t_c},
\hs{2} q F_{ij}\, \rightarrow\, g t_a\, F_{ij\,a}.
\label{7.4.5}
\ee 
As an example we consider the motion of a charged particle on 
the unit sphere $S^2$ supplied with a constant magnetic field. 
The metric on the sphere is defined by the line element
\be 
ds^2 = g_{ij}\, dx^{i} dx^{j} 
 =  d \thg^2 + \sin^2 \thg\, d\vf^2.
\label{7.5}
\ee
The sphere $S^2$ admits a triplet of Killing vectors 
$C^{i} = (C^{\thg}, C^{\vf})$ satisfying the homogeneous form of the 
second Killing equation (\ref{7.4.3}), with $C_{ij}$ and all higher
Killing tensors vanishing:
\be 
\ba{l}
C_{(1)}^{i} = (- \sin \vf,\, - \cot \thg \cos \vf), \\
 \\
C_{(2)}^{i} = (\cos \vf,\, - \cot \thg \sin \vf), \\
 \\
C_{(3)}^{i} = (0,\, 1). 
\ea
\label{7.6}
\ee
These Killing vectors generate three independent rotations on $S^2$.
The magnetic field with constant flux $B$ is described by the field 
strength 
\be
F_{ij}\, dx^{i} \wedge dx^{j} = B \sin \thg\, d\thg \wedge d \vf.
\label{7.7}
\ee
Applying the vectors (\ref{7.6}) in the right-hand side of the
equation for the Killing scalars, we find that they take the form
\be 
C_{(1)} = - qB \sin \thg \cos \vf, \hs{2}
C_{(2)} = - qB \sin \thg \sin \vf, \hs{2}
C_{(3)} = - qB \cos \thg.
\label{7.8}
\ee
Therefore we find as constants of motion the components of the 
gauge-improved angular momentum $J_{(a)}$:
\be 
\ba{l}
J_{(1)} = - \sin \vf\, \Pi_{\thg} - \cot \thg \cos \vf\, \Pi_{\vf} 
 - qB \sin \thg \cos \vf, \\
 \\
J_{(2)} = \cos \vf\, \Pi_{\thg} - \cot \thg \sin \vf\, \Pi_{\vf}
 - qB \sin \thg \sin \vf, \\
 \\
J_{(3)} = \Pi_{\vf} - qB \cos \thg.
\ea
\label{7.9}
\ee
As in eq.\ (\ref{2.22.1}), these constants of motion satisfy the 
$so(3)$ Poisson bracket algebra 
\be 
\left\{ J_{(a)}, J_{(b)} \right\} = \ve_{abc}\, J_{(c)}.
\label{7.10}
\ee
Indeed, the present model is equivalent to a dimensional reduction
of the monopole field from 3-$D$ flat space to the 2-$D$ unit sphere
\ct{plyushchay}. As a result, we also expect the existence of Killing 
tensors. First observe, that
\be 
\sum_a\, J_{(a)}^2 = 2 m H + q^2 B^2,
\label{7.10.1}
\ee
which is a quadratic expression in the momenta with $C_{ij} = g_{ij}$. 
In addition, there two other independent symmetric Killing tensors:
\be 
C_{(1)}^{ij} = \lh \ba{cc} 0 & \cos \vf \\
                           \cos \vf & - 2 \sin \vf \cot \thg \ea \rh, 
\hs{2}
C_{(2)}^{ij} = \lh \ba{cc} 0 & \sin \vf \\
                           \sin \vf & 2 \cos \vf \cot \thg \ea \rh.
\label{7.11}
\ee
Inserting these expressions on the right-hand side of the second
equation (\ref{7.4.3}), and solving this equation, we find the
associated generalized Killing vectors
\be 
\ba{l}
\dsp{ C_{(1)}^{i} = - \frac{qB}{\sin \thg} \lh \cos \thg \sin \thg \cos \vf, 
 ~- \lh \cos^2 \thg - \sin^2 \thg \rh \sin \vf \rh }\\
 \\               
\dsp{ C_{(2)}^{i} = - \frac{qB}{\sin \thg} \lh \cos \thg \sin \thg \sin \vf,
 ~\lh \cos^2 \thg - \sin^2 \thg \rh \cos \vf \rh. }
\ea		  
\label{7.12}
\ee
The associated Killing scalars, the solution of the first equation 
(\ref{7.4.3}), read
\be 
C_{(1)} = q^2 B^2 \sin \thg \cos \thg \sin \vf, \hs{2}
C_{(2)} = - q^2 B^2 \sin \thg \cos \thg \cos \vf.
\label{7.13}
\ee
Combining these results we find the constants of motion
\be 
\ba{lll}
K_{(1)} & = & \dsp{ \cos \vf\, \Pi_{\thg} \Pi_{\vf} 
 - \cot \thg \sin \vf\, \Pi_{\vf}^2 - qB \cos \thg \cos \vf\, \Pi_{\thg} 
 + qB\, \frac{\cos 2 \thg \sin \vf}{\sin \thg}\, \Pi_{\vf} }\\ 
 & & \\
 & & \dsp{ +\, q^2 B^2 \sin \thg \cos \thg \sin \vf, }\\
 & & \\
K_{(2)} & = & \dsp{ \sin \vf\, \Pi_{\thg} \Pi_{\vf} 
 + \cot \thg \cos \vf\, \Pi_{\vf}^2 - qB \cos \thg \sin \vf\, \Pi_{\thg} 
 - qB\, \frac{\cos 2 \thg \cos \vf}{\sin \thg}\, \Pi_{\vf} }\\
 & & \\
 & & \dsp{ -\, q^2 B^2 \sin \thg \cos \thg \cos \vf. }
\ea
\label{7.14}
\ee
Observe, that 
\be 
K_{(1)} = \der_{\vf} K_{(2)}, \hs{2} K_{(2)} = - \der_{\vf} K_{(1)}.
\label{7.15}
\ee
These relations follow, because $\der_{\vf} D_i = D_i \der_{\vf}$. 

\section{Supersymmetry}

Spinning particles whose internal angular momentum is described by
Grassmann co-ordinates $\psi^{i}$ can have Grassmann-odd constants 
of motion, generating transformations in anti-commuting co-ordinates. 
If their bracket closes on the hamiltonian, they generate standard 
supersymmetries. In the case of charged particles in an external 
gauge field, the standard supercharge takes the form
\be 
\Og = \Pi_i \psi^{i},
\label{8.1}
\ee
whilst the non-zero covariant brackets are 
\be 
\left\{ x^{i}, \Pi_j \right\} = \del^{i}_j, \hs{2}
\left\{ \Pi_i, \Pi_j \right\} = q F_{ij}, \hs{2}
\left\{ \psi^{i}, \psi^{j} \right\} = -i \del^{ij}.
\label{8.2}
\ee
in the abelian case, with appropriate modifications in the 
non-abelian generalization. It follows, that the internal spin
rotations are generated by the bilinears
\be 
s_i = - \frac{i}{2}\, \ve_{ijk}\, \psi^{j} \psi^{k}, \hs{2}
\left\{ s_i, s_j \right\} = \ve_{ijk} s_k.
\label{8.3}
\ee
The hamiltonian in flat space reads
\be 
H = \frac{1}{2m}\, \bfPi^2 - \frac{q}{m}\, \bfB \cdot \bfs,
\label{8.4}
\ee
and satisfies the supersymmetric bracket relation
\be 
\left\{ \Og, \Og \right\} = -2mi\, H.
\label{8.5}
\ee
In such a theory any dynamical quantity of which the bracket with 
the supercharge vanishes, is automatically a constant of motion:
\be 
\left\{ Q, \Og \right\} = 0 \hs{1} \Rightarrow \hs{1}
\left\{ Q, H \right\} = 0,
\label{8.6}
\ee
owing to the Jacobi identity for the brackets. The reverse does
not hold in general. Hence the class of superinvariants is a 
subclass of the constants of motion. For these superinvariants
one can derive another more restrictive hierarchy of conditions 
which are sufficient, though in general not necessary, to obtain
solutions of equations (\ref{2.12}) or their appropriate 
generalizations (\ref{4.4}) or (\ref{7.4.3}). These equations were
derived in \ct{grh}, hence it is not necessary to elaborate on them 
in detail. The generating equation is
\be 
-i \psi^{i} \lh \nabla_i Q - q F_{ij} \dd{Q}{\Pi_j} \rh 
 + \Pi_i \dd{Q}{\psi_i} = 0,
\label{8.7}
\ee
obtained by writing out the bracket $\left\{ \Og, Q \right\}$. 
The hierarchy of square roots of the extended Killing equations
is obtained by expanding $Q$ in a series in the momenta $\Pi_i$,
and each coefficient $C_{i_1 ... i_n}(x, \psi)$ in a (finite)
polynomial in the Grassmann variables $\psi^{i}$. In some cases 
of physical interest the superinvariants do not only include known 
constants of motion, such as the angular momentum in the case of a 
monopole field, but also new Grassmann-odd invariants. The coefficients 
in the expressions for such conserved odd charges are generalizations 
of the so-called Killing-Yano tensors, rather than the Killing tensors 
proper. In the case of the magnetic monopole such an anti-commuting 
constant of motion is the non-standard supercharge 
\ct{dejonghe-mcf-peeters-vh}
\be 
\tilde{Q} = \ve_{ijk} \lh x^{i} \Pi^{j} \psi^{k} - \frac{i}{3}\, 
 \psi^{i} \psi^{j} \psi^{k} \rh,
\label{8.8}
\ee 
the bracket of which with itself closes on the square of the angular 
momentum, rather than on the hamiltonian:
\be
\left\{ \tilde{Q}, \tilde{Q} \right\} = - i \lh \bfJ^2 
 - 2 gq\, \frac{\bfr \cdot \bfs}{r} \rh,
\label{8.9}
\ee
with $\bfJ = \bfr \times \bfPi + \bfs$. A generalization for the 
non-abelian monopole was constructed in \ct{mcf-mountain}.
      
\section{Summary and discussion}

In this paper we have developed an algorithm to construct all constants 
of motion for conservative dynamical systems. The method, based on 
extensions of the Killing equations in differential geometry, works in 
the presence of gauge interactions and in non-flat geometries as well. 
It brings out in particular the importance of tuning the symmetries of 
the external fields with those of the geometry of the configuration 
space. The method has been illustrated with several examples, in 
particular monopole-type solutions in abelian and non-abelian gauge 
theories. We have restricted ourselves to classical dynamical systems, 
but the use of a bracket formulation on phase space allows easy 
translation --modulo operator ordering-- to the case of quantum systems.
Also, we have not included spin degrees of freedom. Extending the 
particle models with Grassmann variables to describe fermions 
opens the possibility to include supersymmetries in this framework. 

Apart from the generic importance of constructing constants of motion,
the classification of all dynamical variables commuting with the 
hamiltonian is a starting point for the procedure of hamiltonian 
reduction. This procedure provides an elegant way of constructing 
non-trivial integrable models; for a recent review see \ct{nersessian}. 
This technique was applied to derive $N = 4$ supersymmetric mechanics
in a monopole background in \ct{bel-ner-yer}. The connection with Killing 
vectors and tensors and their generalizations discussed here provides a 
geometrical configuration-space description of this reduction process.

\end{document}